\documentclass[12pt,oneside]{article}
\usepackage{amsmath}
\usepackage{amsfonts,amssymb,graphicx}
\def\be{\begin{equation}}
\def\ee{\end{equation}}
\def\bea{\begin{eqnarray}}
\def\eea{\end{eqnarray}}
\def\<{\langle}
\def\>{\rangle}
\def\~{\tilde}
\def\s{\sigma}

\def\a{\alpha}
\def\b{\beta}

\newcommand{\R}{\Bbb R}
\newcommand{\E}{\Bbb E}

\newcommand{\N}{\Bbb N}

\newcommand{\Z}{\Bbb Z}

\newcommand{\av}[1]{\mathbb E\left[#1\right]}

\newtheorem{theorem}{Theorem}[section]
\newtheorem{remark}{Remark}
\newtheorem{proposition}[theorem]{Proposition}
\newtheorem{lemma}[theorem]{Lemma}
\newtheorem{definition}[theorem]{Definition}
\newtheorem{corollary}[theorem]{Corollary}

\newenvironment{proof}{{\bf Proof:}}{\hfill$\square$\vskip.5cm}
\renewcommand{\P}{\mathbb{P}}

\begin{document}
\begin{center}
{\bf\sc\Large
Variational Bounds for the\\
Generalized Random Energy Model\\}
\vspace{1cm}
{Cristian Giardin\`a}\\
\vspace{.5cm}
{\small EURANDOM} \\
{\small P.O. Box 513 - 5600 MB Eindhoven, The Netherlands}\\
{\small {e-mail: {\em giardina@eurandom.tue.nl}}}\\
\vskip 0.5truecm
\vskip 0.5truecm
{Shannon Starr}\\
\vspace{.5cm}
{\small UCLA Mathematics Department} \\
{\small Box 951555, Los Angeles, CA 90095-1555}\\
{\small {e-mail: {\em sstarr@math.ucla.edu}}}\\
\vskip 1truecm

\end{center}
\vskip 1truecm
\begin{abstract}
\noindent
We compute the pressure of the random energy model (REM) and
generalized random energy model
(GREM) by establishing variational upper and lower bounds.
For the upper bound, we generalize Guerra's ``broken replica symmetry bounds",
and identify the random probability cascade as the appropriate random
overlap structure for the model.
For the REM
the lower bound is obtained,
in the high temperature regime
using Talagrand's concentration
of measure inequality,
and in the low temperature regime using convexity and the high temperature formula.
The lower bound for the GREM follows from the lower bound for the REM by
induction.
While the argument for the lower bound is fairly standard, our proof of the upper
bound is new.

\end{abstract}
\newpage\noindent

\section{Introduction}

While the analysis and properties of finite range spin glass systems
(like the Edwards-Anderson model) is still a very
debated issue, even in the physics community, in recent years there has been a large progress in
the mathematical understanding of mean-field models \cite{MPV}. This advance was
triggered by the introduction of a ``quadratic interpolation'' technique,
pioneered in \cite{GT} to establish the existence of thermodynamic
limit for the Sherrington-Kirkpatrick (SK) model \cite{SK} and further developed
by Guerra in \cite{G} to prove an upper bound for the pressure which
coincides with the Parisi replica symmetry breaking solution of the model.
Motivated by the cavity picture, this bound was generalized to a variational
bound by the introduction of Random Overlap Structures (ROSt) and associated random weights in \cite{AiSS}.
Guerra's bound is recovered when the weights are chosen from the inhomogeneous Poisson point
process studied by Ruelle \cite{R} and the ROSt is a hierarchical one.
Recently Talagrand \cite{T2} was able to prove that Guerra's bound is optimal,
by showing that the correction term in the bound goes to zero in the thermodynamic limit,
thus establishing the rigorous validity of the Parisi solution.

In this paper we solve the Generalized Random Energy Model (GREM)
\cite{D,DG}
by a simple analysis that partially follows the one developed for
the SK model. We first obtain a variational upper bound for the
pressure by the definiton of the appropriate auxiliary system (i.e., ROSt)
for the model. As a guideline in this step we use the basic
covariance inequality that was identified in \cite{CDGG}.
To show that the upper bound is optimal when the ROSt
random weights are chosen according to the Poisson-Dirichlet
point process we use a different strategy than the one developed for
the SK model. Indeed the corresponding lower bound is easily obtained
from a complete control of the high temperature region and convexity of
the pressure. To study the high temperature region we propose a
new induction argument, which starting from infinite temperature
covers all the temperature values up to the critical one.

Despite the recent progress, there is not
yet a direct proof of the most prominent property of the Parisi solution,
namely {\em ultrametricity}. A distance $d$ in a metric space is said to be
ultrametric if the standard triangular inequality is replaced by the much stronger
inequality $d(x,z) \leq \max \{d(x,y),d(y,z)\}$. Equivalently, in an ultrametric
space all triangles are equilateral or isosceles with longer equal sides.
The Generalized Random Energy Model (GREM), introduced by Derrida \cite{D}
as a model that possesses replica symmetry breaking, is ultrametric by
construction, since its Hamiltonian is defined as the
sum of independent Gaussian random variables positioned on the branches of a
hierarchical tree. As a first step in the direction of achieving
a proof of ultrametricity for mean-field models, generalized
non-hierarchical models have been considered in \cite{BK},
where it has been shown that they exhibit  GREM-like behaviour.
The analysis of the present paper, which sets the GREM model
into the general variational scheme developed for the SK model,
could be helpful in further studying non-hierarchical models.

The paper is organized as follow. In Section 2 we treat the basic
case of the Random Energy Model (REM), where the energy levels are independent
Gaussian random variables. This is a necessary warm-up since the GREM
model will be treated as a nested succession of REM-like systems.
The full analysis is presented in Section 3. In the Appendix we
recall some useful concentration of measure estimates.

\vspace{0.1cm}
{\bf Acknowledgements:} We thank A.~Bovier and
A.C.D.~Van Enter for helpful comments and suggestions to improve
the presentation of the results. We also thank A.~Bianchi for early
discussions, and an anonymous referee for numerous helpful improvements.
C.G. acknowledges P.~Contucci and
S.~Graffi for their encouragement in this work.
His research in
conducted under financial support of NWO-project 613000435.

\section{REM}

The random energy model (REM)
is a statistical mechanical model, where the energy levels are independent
and identically distributed Gaussian random
variables.
More precisely, for a system of size $N$,
the Hamiltonian is a Gaussian centered family
with covariance matrix
\be
\label{cov}
 C_H(\s,\s') := \E[H_N(\s)H_N(\s')] = \frac{N}{2}\, \delta(\s,\s')
\ee
where $\s,\s'  \in \{+1,-1\}^N$ are vectors, whose components are
Ising spin variables.
The dependence on $N$ in (\ref{cov}) is
such that thermodynamic observables (energy, free energy, etc.) are extensive
in the volume, while the factor $1/2$ is included as a matter of convention.
    In this paper we denote by $X$ a standard Gaussian,
    $\av{X} = 0$, $\av{X^2} = 1$, thus an explicit representation of the Hamiltonian is
    \be
    \label{eq:HFunE}
    H_N(\s) = \sqrt{\frac{N}{2}}\,X(\s)
    \ee
    where $\{X(\s)\}_{\s\in\Sigma_N}$ are $2^N$ i.i.d.\ copies of the random variable $X$.
An equivalent representation, more in the spirit of statistical mechanics,
is to consider a ``lattice" $\Lambda$ with
$N=|\Lambda|$  sites and a random Hamiltonian
\be
\label{eq:HFunJ}
H_{\Lambda}(\s) = \sqrt{\frac{N}{2^{N+1}}}\sum_{X \subseteq \,\Lambda} J_X \s_X
\ee
defined on the spin configurations $\sigma : \Lambda \to \{+1,-1\}$.
In (\ref{eq:HFunJ}) the $J_X$'s are a family of i.i.d.\ Gaussian random variables,
with $\E[J_X^2] = 1$, and $\s_X := \prod_{i\in X} \s_i$ for each $X \subseteq \Lambda$.
Henceforth, all Gaussian random variables will be understood to have expectation equal to 0.

The main quantity we are going to study
is the quenched {\em pressure}.
We denote by $\Sigma_N = \{-1,+1\}^{N}$ the space of all possible
spin configurations. For a finite system we define the
{\em random} partition function
\begin{equation}
Z_N(\beta)\, =\, \sum_{\sigma \in \Sigma_N} e^{-\beta H_N(\sigma)}\, ,
\end{equation}
and the {\em quenched} pressure
\be
\label{press}
P_N(\b)\, =\,  \E\bigg[\frac{1}{N} \ln Z_N(\beta)\bigg]\, .
\ee
We will be interested in the thermodynamic limit
\begin{equation}
\label{tdlim-press}
P(\b)\, =\, \lim_{N \to \infty} P_N(\b)
\end{equation}
%
In the following we will sometimes drop the $N-$dependence in the Hamiltonian $H(\s)$ in order to alleviate
notation. We will also introduce additional randomness by considering an auxiliary
system which is coupled to the Hamiltonian. We will denote by $\av{\cdot}$
the expectation with respect to all random variables involved.

\subsection{Upper bound}

We start by recalling the quadratic interpolation technique.
\begin{lemma}
\label{QI}
Let $H(\s)$ be a Gaussian family, indexed by $\s \in \Sigma_N$, with covariance $C_H(\s;\s')$.
Let $\alpha \in \mathcal{A}$ be an index ranging over the set $\mathcal{A}$,
and let $K(\a)$ and $V(\s,\a)$ be Gaussian
random variables, independent of $H(\s)$ and of each other, with covariances $C_K(\a;\a')$ and $C_V(\s,\a;\s',\a')$,
respectively.
Suppose that
\be
\label{cov-ineq}
C_H(\s;\s') + C_K(\a;\a') \ge C_V(\s, \a; \s', \a')
\ee
for all $\s,\s'\in\Sigma_N$ and $\a,\a'\in\mathcal{A}$,
and suppose that
\be
\label{cov-equal}
C_H(\s;\s) + C_K(\a;\a) = C_V(\s, \a; \s, \a)
\ee
for every $\s \in \Sigma_N$ and $\a \in \mathcal{A}$.
Moreover, suppose that there is a random weight $w : \mathcal{A} \to [0,\infty)$
such that, almost surely, $\sum_{\alpha \in \mathcal{A}} w(\alpha)$
is strictly positive and finite.
Then,
\begin{equation}
\label{upper-ineq}
\begin{split}
P_N(\b)\,
&\le\, \E\Bigg[ \frac{1}{N} \ln \frac{\sum_{\s,\a} w(\a) e^{-\b V(\s,\a)}}
{\sum_{\a} w(\a) e^{-\b K(\a)}}\Bigg]\\
&=\, \E\Big[\frac{1}{N} \ln \sum\nolimits_{\sigma,\alpha} w(\alpha) e^{-\beta V(\sigma,\alpha)}\Big]
- \E\Big[\frac{1}{N} \ln \sum\nolimits_{\alpha} w(\alpha) e^{-\beta K(\alpha)}\Big]\, ,
\end{split}
\end{equation}
as long as the right-hand-side is well-defined (i.e., not $\infty - \infty$).
\end{lemma}

\begin{proof}
We refer to \cite{GT,AiSS} for full details. Here we only recall the
basic idea. For $t\in[0,1]$ define an interpolating Hamiltonian
\be
\tilde{H}(\sigma,\alpha;t)\, =\, \sqrt{1-t} \;[H(\sigma) + K(\alpha)] + \sqrt{t} \; V(\sigma,\alpha)\, .
\ee
and an associated random partition function
\be
Z_{N,\,t}(\b) = \sum_{\s,\a} w(\a) e^{-\beta \tilde{H}(\sigma,\alpha \,; t)} \;.
\ee
Let $\Omega_{N,\beta,t}$ denote expectation with respect to the multiple-replica product
measure, where the weight for a configuration $(\s,\a)$ of a generic copy is given by
Gibbs measure associated to $\tilde H(\s,\a)$ times a generic weights $w(\a)$.
In particular, for a function $f(\s,\a,\s',\a')$ of two replicas,
one has
$$
\Omega_{N,\beta,t}\{f(\sigma,\alpha ;\sigma',\alpha')\}\,
=\, \sum\nolimits_{\sigma,\alpha} \sum\nolimits_{\sigma',\alpha'} w(\alpha) w(\alpha') \;
\frac{e^{-\beta [ \tilde{H}(\sigma,\alpha;t) + \tilde{H}(\sigma',\alpha';t) ]}}{Z_{N,\,t}^2}
f(\sigma,\alpha;\sigma',\alpha')\, .
$$
Then one actually has (because of the equality along the diagonal Eq. (\ref{cov-equal}))
\be
\label{sum-rule}
\begin{split}
&\hspace{-30pt}
\frac{1}{N} \;\E\Big[\ln \sum\nolimits_{\sigma,\alpha} w(\alpha) e^{-\beta V(\sigma,\alpha)}\Big]
- \frac{1}{N}\; \E\Big[\ln \sum\nolimits_{\alpha} w(\alpha) e^{-\beta K(\alpha)}\Big]
- P_N(\beta)\\
&=\, \frac{\beta^2}{2} \int_0^1 \E[\Omega_{N,\beta,t}\{C_H(\sigma;\sigma') + C_K(\alpha;\alpha') - C_V(\s,\alpha;\s',\alpha')\}]\, dt\, .
\end{split}
\ee
This is proved by differentiating the quantity
$$
\E[N^{-1} \ln \sum\nolimits_{\alpha,\sigma} w(\alpha) e^{-\beta \tilde{H}_N(\sigma,\alpha;t)}]\, ,
$$
with respect to $t$ and using the generalized Wick's rule.
Because of (\ref{cov-ineq}), the right hand side of (\ref{sum-rule}) is obviously positive
and equation (\ref{upper-ineq}) follows.
\end{proof}

\begin{remark}
The same basic argument works to bound $\E[F(Z_N(\beta))]$ for other functions
such as $F(z)=z^{a}$.
\end{remark}

\begin{remark}
An identity such as (\ref{sum-rule}) is usually called a sum-rule. The process
$K(\a)$ has to be thought of as a large reservoir which acts on the original
system $H(\s)$ through the interaction $V(\s,\a)$.

\end{remark}
We are going to use the previous lemma to establish an optimal upper
bound for the REM model.
A key element is to choose the correct formula for the random weight $w(\alpha)$.
The correct formula for mean field spin glasses seems to generally be given by Ruelle's random probability cascade.
For the REM, it is given by a single level of that, which is sometimes called the Poisson-Dirichlet
process\footnote{See \cite{PY} but also see \cite{R}. For a rather more abstract version, see \cite{BS}.}.
Let us give a brief description of this (the unnormalized version) in order to facilitate
the following proposition.

Given
$0<m<1$ consider the Poisson point process on $(0,\infty)$ with intensity measure
equal to $m w^{-m-1}\, dw$.
Almost surely, the points can be labelled $\{w_1,w_2,\dots\}$
with $w_1>w_2>\dots>0$.
Moreover,
$\sum_{\alpha=1}^\infty w_\alpha$ is strictly positive and finite, almost surely.
The distribution of $\{w_\alpha\}_{\alpha}$ has a remarkable invariance property:
If $(f_1,f_2,\dots)$ are i.i.d.\ copies of the random variable, $f$,
which are assumed to be independent of $\{w_{\alpha}\}_{\alpha}$,
then (modulo permutations) the distribution of $\{e^{f_\alpha} w_\alpha\}_{\alpha}$
is the same as $\{c w_{\alpha}\}_{\alpha}$
where $c$ is the nonrandom number
$c = (\E[e^{m f}])^{1/m}$.
This is easily proved using the generalized Laplace transform.
(For a proof see, for example, reference \cite{T}, page 481.)

One may note that instead of considering the Poisson point process $\{w_{\alpha}\}_{\alpha}$
with intensity equal to $m w^{-m-1}\, dw$,
one could instead consider $w_{\alpha} = e^{y_{\alpha}/m}$ for some $\{y_{\alpha}\}_{\alpha}$.
Then $-\infty<\dots<y_2<y_1<\infty$ is a Poisson point process on $\R$
with intensity measure $e^{-y}\, dy$ (independent of $m$).
One thinks of $-y_{\alpha}$'s as the free energies.
In this notation $m$ is explicit.
Henceforth $\{y_{\alpha}\}_{\alpha}$ will refer to the point process
just described.

A consequence of the invariance property mentioned above is
\be
\label{invariance}
\sum_{\a=1}^{\infty} e^{y_{\a}/m} \exp (f_{\a})\,
\stackrel{\mathcal{D}}{=}\,
\E\big[\exp (m f)\big]^{1/m} \sum_{\a=1}^{\infty} e^{y_{\a}/m}\, ,
\ee
which will be useful.


\begin{proposition}
\label{prop:upper}
Let $C_H(\s,\s')$ be given by Eq. (\ref{cov}). Choose
$w(\a) = \exp[y(\a)/m]$ for $0<m<1$.
For each $b\ge 1$ let
\be
\label{covK}
C_K(\a,\a') = (b-1) \frac{N}{2}\;\delta(\a,\a')
\ee
\be
\label{covV}
C_V(\s, \a; \s', \a') = b\, \frac{N}{2}\;\delta(\s,\s')\delta(\a,\a').
\ee
Then one obtains the optimal upper bound for the REM,
\be
\label{upper-bound}
P_N(\b) \le \inf_{0 < m < 1} \left[ \frac{1}{4} m \b^2 + \frac{1}{m}\ln 2 \right]
\ee
\end{proposition}
\begin{proof}
We note that Lemma \ref{QI} is applicable because
\be
b \,\delta(\s,\s') \delta(\a,\a') \le \delta(\s,\s') + (b-1) \delta(\a,\a')
\ee
We compute separately the two terms in Eq.(\ref{upper-ineq}).
For the first one, due to Eq. (\ref{covV}), we have
\bea
& & \frac{1}{N}\;\av{\ln \sum_{\s,\;\a}w(\a)\exp [-\b V(\s,\a)]}  = \\
& & \frac{1}{N}\;\av{\ln \sum_{\a}\exp \left[\frac{y(\a)}{m}\right]
\sum_{\s}\exp \left[-\b \sqrt{\frac{b N}{2}} X(\s,\a)\right]} = \\
& & \frac{1}{N}\;\av{\ln \sum_{\a}\exp \left[\frac{y(\a)}{m}\right] Z_{N}(\b\sqrt{b}\,;\a)}
\label{pippo}
\eea
where $Z_{N}(\b\sqrt{b}\,;\a)$ are independent copies (labeled by the $\a$'s) of the random variable
$Z_N(\b\sqrt{b}) = \sum_{\s} \exp \left[-\b\sqrt{\frac{bN}{2}} X(\s)\right ]$.
By applying the invariance property of Eq.(\ref{invariance}) to Eq.(\ref{pippo}) with $\exp[f_{\a}] = Z_N(\b\sqrt{b};\a)$ we obtain
\bea
& & \frac{1}{N}\;\av{\ln \sum_{\s,\;\a}w(\a)\exp [-\b V(\s,\a)]}  = \\
& & \frac{1}{N}\;\av{\ln \sum_{\a}\exp \left[\frac{y(\a)}{m}\right]} +
\frac{1}{mN}\ln \av{Z_N^m(\b\sqrt{b})}.
\label{carolina}
\eea
Then we consider the second term in Eq.(\ref{upper-ineq}). Taking into account
the choice (\ref{covK}) we have
\bea
& & \frac{1}{N}\;\av{\ln {\sum_{\a}\exp \left[\frac{y(\a)}{m}\right]\exp [-\b K(\a)]}} = \\
& & \frac{1}{N}\;\av{\ln {\sum_{\a}\exp \left[\frac{y(\a)}{m}\right]
\exp \left[-\b \sqrt{\frac{(b-1)N}{2}} X_{\a}\right]}}.
\eea
Using again the invariance property Eq.(\ref{invariance}) with
$\exp[f_{\a}] = \exp \left[-\b \sqrt{\frac{(b-1)N}{2}} X_{\a}\right]$
and computing the average we obtain
\bea
& & \frac{1}{N}\;\av{\ln {\sum_{\a}\exp \left[\frac{y(\a)}{m}\right]\exp [-\b K(\a)]}} = \\
& & \frac{1}{N}\;\av{\ln \sum_{\a}\exp \left[\frac{y(\a)}{m}\right]} +
\b^2 \frac{(b-1) m}{4}\;.
\label{pluto}
\eea
Putting together Eq.(\ref{carolina}) and (\ref{pluto}) we obtain:
\be
\label{inter}
P_N(\b) \; \le \;
\frac{1}{mN}\ln \av{Z_N^m(\b\sqrt{b})} - \b^2 \frac{(b-1) m}{4} \;.
\ee
Now we use the simple fact that
\be
\label{ciao}
Z_N^m(\b\sqrt{b}) \leq Z_N(m\b\sqrt{b}).
\ee
This is a general fact in statistical mechanics: since the entropy is positive by definition,
the free energy is increasing in $\beta$. Indeed, considering
$f_N(\beta) = - \frac{1}{N\beta} \ln(Z_N(\beta))$,
the {\it random} free energy, one immediately checks that
$f'_N(\beta) = \frac{1}{\beta}\,(u_N(\beta)-f_N(\beta)) =
\frac{1}{\beta^2} \, s_N(\beta) \geq 0$, where $u_N(\beta)$ is the {\it random} internal energy,
and $s_N(\beta)$ is the {\it random} entropy.
Therefore, for any $0<m\leq 1$, we have:
\be
- \frac{1}{Nm\beta} \ln(Z_N(m\beta)) \leq - \frac{1}{N\beta} \ln(Z_N(\beta))\, ,
\ee
which is equivalent to (\ref{ciao}) when we replace $\beta$ by $\beta \sqrt{b}$.
By inserting Eq.(\ref{ciao}) into Eq.(\ref{inter}) it is now easy to compute the
expectation $\av{Z_N(m\b\sqrt{b})}$ and we arrive at the upper bound
\be
\label{easycalc}
P_N(\b) \le \frac{1}{4} m \b^2 + \frac{1}{m}\ln 2 \;.
\ee
Note that there is not anymore dependence on $b$ in the bound.
Finally the optimal bound is obtained by minimization in $m$
which yields the (\ref{upper-bound}) and completes the proof.
\end{proof}

\begin{remark}
\label{rem:Jensen}
It is interesting to note that for the REM (and for the REM only) the main result of quadratic interpolation can be viewed as an implementation of Jensen's inequality:
\begin{equation}
\begin{split}
\E[\ln(Z_N(\beta))]\,
    =\, \frac{1}{m} \, \E[\ln Z_N^m(\beta)]\,
    \leq\, \frac{1}{m} \ln \E[Z_N^m(\beta)]\, .
\end{split}
\end{equation}
This holds for general $m>0$, but one needs $m\leq 1$ to apply (\ref{ciao}).
If one uses the ``sum-rule" then one can get an explicit form for the {\it error} coming from Jensen's inequality in this case.
\end{remark}


\begin{remark}
\label{rem:actualvalue}
The inequality of (\ref{upper-bound}) takes two different forms depending on whether $\beta$ is greater than or less than $\beta_c := 2 \sqrt{\ln(2)}$. For $\beta > \beta_c$, the right-hand-side of (\ref{upper-bound}) is optimized at $m=\beta_c/\beta$. For $\beta<\beta_c$ the infimum over $0<m<1$ is attained by a limit $m\to 1$.
Therefore, one has
\begin{equation}
P_N(\beta)\, \leq\, \mathcal{Q}(\beta)\, ,
\end{equation}
where
\begin{equation}
\label{Q-rem}
\mathcal{Q}(\beta)\, =\, \left\{\begin{array}{ll}  \frac{1}{4} \beta^2 + \ln(2) & \textrm{ for } \beta < \beta_c\\
\beta \sqrt{\ln(2)} & \textrm{ for } \beta \geq \beta_c
\end{array} \right\} \, =\, \left\{\begin{array}{ll}   \frac{1}{4} (\beta^2 + \beta_c^2) & \textrm{ for } \beta < \beta_c\\
 \frac{1}{2} \beta \beta_c & \textrm{ for } \beta \geq \beta_c
\end{array} \right\}
\end{equation}
Note that the two pieces match at $\beta=\beta_c$.
\end{remark}


\subsection{Lower bound}
\label{subsec:REMlb}

The main new result in this paper is the adaptation of the quadratic
interpolation method to obtain an asymptotically sharp upper bound on $P_N(\beta)$,
as we just considered above.
One should view the result of the previous section as an analogue, for the REM,
of Guerra's bounds for the SK model in \cite{G}.
On the other hand, for the REM, unlike for the SK model, there are easy proofs
of the same lower bound in the $N\to \infty$ limit.
The exact formula for the pressure of the random energy model is well-known.
Derrida calculated it when he introduced the model in \cite{D} while
a mathematically rigorous version of his argument
is included in \cite{B,OP}.
There are also proofs which rely more on large-deviation
{\it theory}, such as \cite{Eis,DW}
\footnote{In \cite{Eis}, there is also a conjecture for the more general quantity
$N^{-1} \ln E[Z_N(\beta)^a]$ for $a \in \R$. This is substantiated in the preliminary
section of \cite{T3}.}.
We also encourage the reader to see the more recent, deep analysis of
\cite{BK1,BK2}.
(There is much interest in the REM as far as the statistics of energy levels
is concerned because in the bulk there is a kind of universality. See, for instance,
\cite{BK3}.)

We will also present a proof of the lower bound.
This is included primarily for completeness, for {\it nonexperts}.
However, let us digress briefly to justify this for the experts:
there is a great desire to obtain a purely variational proof of the lower bound
for the the SK model.
In light of that, it seemed worthwhile to explore how `variational'
the proof of the lower bound really is, for the REM.
(Of course, it is not as variational as one would like).

Since $P_N(\beta) \leq \mathcal{Q}(\beta)$ for all $N$, it follows that $P(\beta) \leq \mathcal{Q}(\beta)$ in the limit $N\rightarrow \infty$.
We want to show the opposite is also true, to establish that $P(\beta)=\mathcal{Q}(\beta)$ for all $\beta\geq 0$.
The key to obtaining the lower bound is to understand the high temperature region, $\beta < \beta_c$.
\begin{proposition}
\label{prop:LowerBd}
\begin{equation}
\label{ineq:Lower}
P(\beta)\, =\,  \mathcal{Q}(\beta)\,\quad \textrm{ for }\quad \beta\leq \beta_c\, .
\end{equation}
\end{proposition}

The proof of this result is provided in the next subsection, while here we stress
its consequences.
Stated otherwise, the upper bound of Proposition \ref{prop:upper} saturates in the $N\to\infty$ limit, at least when $\beta \leq \beta_c$.
It is a remarkable fact that this high-temperature result gives the
sharpness of the upper bound also
in the low-temperature region as follows.
\begin{corollary}
\label{cor:complement}
For $\b\in[0,+\infty)$
\end{corollary}
\begin{equation}
P(\beta)\, =\, \mathcal{Q}(\beta) .
\end{equation}
\begin{proof}
It is a basic fact, easily seen from the definition (\ref{press}),
that $P_N(\beta)$ is convex in $\beta$ for each $N$.
Therefore, the limiting function $P(\beta)$ is also convex.
Hence, for any $\beta_0$ and any $\beta\geq \beta_0$, we have
\be
P(\beta) \ge P(\beta_0) + (\beta - \beta_0) DP(\beta_0)\, ,
\ee
where $D$ is any convex combination of the left-handed and right-handed derivatives, which we denote $D_-$ and $D_+$, respectively.
We now take $\beta_0 \uparrow \beta_c$.
Since we know from Proposition \ref{prop:LowerBd}
that $P(\beta)=\mathcal{Q}(\beta)$ for $\beta< \beta_c$, we easily calculate
$\lim_{\beta_0 \uparrow \beta_c} P\big(\beta_0) = 2 \ln(2)$ while $\lim_{\beta_0 \uparrow \beta_c} D_-P(\beta_0) = \lim_{\beta_0 \uparrow \beta_c} \mathcal{Q}'(\beta_0) = \sqrt{\ln (2)}$.
Putting this together completes the proof.
\end{proof}

\subsection{High temperature region}

The proof of Proposition  \ref{prop:LowerBd} will be obtained throught a sequence
of lemmata.
The crux of the argument is standard.
For example, see \cite{T}, Proposition 1.1.5.
(there is another approach in \cite{B}, Theorem 9.1.2, called the ``truncated second moment method'').
We start with the following result, which it is another variational calculation.
%
%
\begin{lemma}
\label{lem:double-Z-estimate}
Let $\Omega_{\beta,N}$ refer to the (expectation associated to the) random probability measure on $\Sigma_N \times \Sigma_N$
specified by
\begin{equation*}
\Omega_{\beta,N}\big\{f(\sigma,\sigma')\big\}\,
:=\, Z_N^{-2}(\beta) \sum\nolimits_{\sigma, \sigma' \in \Sigma_N} e^{-\beta H(\sigma)} e^{-\beta H(\sigma')} f(\sigma,\sigma')\, .
\end{equation*}
For $0\leq \beta\leq \beta_c$ we have
\begin{equation}
\label{ineq:double-Z-estimate}
\frac{1}{N} \ln \E\big[Z_N(\beta) \Omega_{\beta,N}\big\{\delta(\sigma,\sigma')\big\}\big]\, \leq\, \frac{\beta \beta_c}{2}\, .
\end{equation}
\end{lemma}
\begin{proof}
Note that
\be
\Omega_{\beta,N}\big\{\delta(\sigma,\sigma')\big\}\,
=\, \frac{\sum_{\sigma, \sigma' \in \Sigma_N} e^{-\beta H(\sigma)} e^{-\beta H(\sigma')} \delta(\sigma,\sigma')}{Z^2_N(\beta)}\,
=\,
\frac{Z_N(2\beta)}{Z^2_N(\beta)}\, .
\ee
By (\ref{ciao}), we know that for $0<m< 1$
\be
Z_N(2\beta)\, \leq\, Z^{1/m}_N(2m\beta)\,  .
\ee
Using H\"older's inequality with $p=1/m$, $q=1/(1-m)$ and $\frac{1}{p} + \frac{1}{q} =1$
\be
Z_N(2m\beta)\, \leq\, Z^m_N(\beta)\, Z_N^{1-m}\Big(\frac{m}{1-m} \beta\Big)\, .
\ee
Therefore,
$$
Z_N(\beta) \Omega_{\beta,N}\big\{\delta(\sigma,\sigma')\big\}\,
=\, \frac{Z_N(2\beta)}{Z_N(\beta)}\,
\leq\, \frac{Z_N^{1/m}(2m\beta)}{Z_N(\beta)}\,
\leq\, Z_N^{(1-m)/m}\Big(\frac{m}{1-m} \beta\Big)\, .
$$
Since $m/(1-m)$ can take any positive value as $m$ ranges over $(0,1)$ this means
\begin{equation}
\label{eq:helper1}
Z_N(\beta) \Omega_{\beta,N}\big\{\delta(\sigma,\sigma')\big\}\,
\leq\, Z_N^{1/r}(r \beta)
\end{equation}
for every $r \in [0,+\infty)$.
Moreover, for $r\geq 1$ we can use Jensen's inequality to obtain
\begin{equation}
\label{eq:helper2}
\E[Z_N^{1/r}(r \beta)]\, \leq\, \big(\E[Z_N(r\beta)]\big)^{1/r}\, =\, \exp\Big(N\Big[\frac{\beta_c^2}{4r} + \frac{r \beta^2}{4}\Big]\Big)\, .
\end{equation}
It is easy to see that the optimal value is $r=\beta_c/\beta$, which does satisfy the constraint $r\geq 1$ because
of the hypothesis $\beta \leq \beta_c$.
Choosing this $r$ and putting (\ref{eq:helper1}) and (\ref{eq:helper2}) together yields
(\ref{ineq:double-Z-estimate}).
\end{proof}
A second useful estimate is the following concentration of measure property.
\begin{lemma}
\label{lem:COM}
For any $\beta$, and any $t>0$,
$$
\P\big\{|N^{-1} \ln Z_N(\beta) - P_N(\beta)|\geq \beta t\big\}\, \leq\, 2 e^{-Nt^2/2}\, .
$$
\end{lemma}
The analogous result for the SK model is Corollary 2.2.5 of \cite{T}.
For completeness we will give a proof of Lemma \ref{lem:COM} in Appendix  A and, particularly, show that a straightforward generalization of Talagrand's proof applies equally well to
all models that satisfy thermodynamic stability \cite{CGi,CGr}.

The proof of Proposition  \ref{prop:LowerBd} will essentially follow from the next result,
which we prove first.
\begin{lemma}
\label{prop:overlap}
For any $0\leq \beta < \beta_c$,
\begin{equation}
\label{ineq:overlap}
\limsup_{N \to \infty}
 \sup_{0\leq \beta'\leq \beta}
\frac{1}{N} \ln \E\big[\Omega_{\beta',N}\big\{\delta(\sigma,\sigma')\big\}\big]\,
<\, 0\, .
\end{equation}
\end{lemma}
\begin{proof}
The proof will obtained by induction. Let us define the succession of temperatures
given by $\b_0 = 0$, $\b_{n+1}=g(\b_n)$ for $n\in \N$, where $g$ is a definite
function.
As we will see, we can choose
\be
\label{map}
g(\beta) =  \beta + a \beta_c [1-(\beta/\beta_c)]^2
\quad \mbox{for any} \quad 0<a<1/2.
\ee
and it will follow that $\beta_n \uparrow \beta_c$ as $n \to \infty$.

We first note that (\ref{ineq:overlap}) is true for $\b=\b_0 = 0$,
because  one has $\E\big[\Omega_{0,N}\big\{\delta(\sigma,\sigma')\big\}\big]\, =\, 2^{-N}$.
For the induction step, we will prove that if (\ref{ineq:overlap}) is true for $\beta\in [0,\b_n]$,
then it is also true for every $\beta \in [0,\b_{n+1}]$.
Then, since $\beta_n \uparrow \beta_c$
the statement of Lemma \ref{prop:overlap} follows.

To prove the induction step, we first observe that
\begin{equation}
\label{correction}
\frac{d}{d\beta} P_N(\beta)\,
=\, \frac{\beta}{2} \left(1 -  \E\big[\Omega_{N,\beta}\big\{\delta(\sigma,\sigma')\big\}\big]\right)\,
\geq 0 \,.
\end{equation}
Indeed, this is a  simple calculation using the generalized Wick's rule.
Since $Z_N(0)=2^N$ (deterministically) we have $P_N(0) = \ln(2)$. Then the sum rule
\begin{equation}
\label{sumrule-beta}
P_N(\beta)\, =\, \ln(2) +  \frac{\beta^2}{4} - \int_0^{\beta} \frac{\beta'}{2} \
\E\big[\Omega_{N,\beta'}\big\{\delta(\sigma,\sigma')\big\}\big]\, d\beta'
\end{equation}
follows.

Suppose now that (\ref{ineq:overlap}) is true
for $\beta\in [0,\b_n]$.
Consider a generic $\beta > \b_n$, not necessarily smaller than $\beta_{n+1}$, and for $t>0$
let $A_N(\beta,t)$ be the event in Lemma \ref{lem:COM}:
\begin{equation*}
A_N(\beta,t)\, =\, \big\{|N^{-1} \ln Z_N(\beta) - P_N(\beta)|\geq \beta t\big\}\, .
\end{equation*}
On $A_N(\beta,t)^c$ we have $Z_N(\beta) \geq e^{N [P_N(\beta) - \beta t]}$.
Let us employ the following shorthand: given a set $A$, let $I_A$ denote the indicator of $A$ and let us denote $\E[I_A X]$ by $\E[X,A]$ for each and every  random variable $X$.
Then we conclude that
\begin{align*}
\E\big[  \Omega_{\beta,N}\big\{\delta(\sigma,\sigma')\big\},\, A_N(\beta,t)^c\big]\,
&\leq\, \E\Big[  \frac{Z_N(\beta)}{e^{N [P_N(\beta) - \beta t]}}
\Omega_{\beta,N}\big\{\delta(\sigma,\sigma')\big\},\, A_N(\beta,t)^c \Big] \nonumber \\
&\leq\, e^{-N [P_N(\beta) - \beta t]}  \E\big[Z_N(\beta) \Omega_{N,\beta}\big\{\delta(\sigma,\sigma')\big\} \big] \, .
\end{align*}
Therefore
\begin{equation*}
\frac{1}{N} \ln\E\big[  \Omega_{N,\beta}\big\{\delta(\sigma,\sigma')\big\},\, A_N(\beta,t)^c\big]\,
\leq\, - P_N(\beta) + \frac{1}{2} \beta \beta_c + \beta t
\end{equation*}
follows from Lemma \ref{lem:double-Z-estimate}.

Since we assumed $\beta>\beta_n$,
a lower bound for $P_N(\beta)$ is given by
$P_N(\beta_n)$
and a lower bound for $P_N(\beta_n)$ is given by $\frac{1}{4} [\beta_c^2 + \beta_n^2] - o(1)$,
using Eq. (\ref{sumrule-beta}) and the induction hypothesis,
where $o(1)$ represents a quantity whose limit is $0$ when $N\to \infty$.
One thing which is important is that while $o(1)$ does depend on $\beta_n$,
it is independent of $\beta>\beta_n$.
Therefore,
\begin{equation}
\label{eq:Helper1}
\frac{1}{N} \ln\E\big[  \Omega_{\beta,N}\big\{\delta(\sigma,\sigma')\big\},\, A_N(\beta,t)^c\big]\,
\leq\, - \frac{1}{4} (\beta_c - \beta_n)^2 + \frac{1}{2} \beta_c [\beta-\beta_n] + \beta t - o(1)\, .
\end{equation}
On the other hand, one always has $0\leq \Omega_{\beta,N}\big\{\delta(\sigma,\sigma')\big\}\leq 1$. Hence,
\begin{equation*}
\E\big[  \Omega_{\beta,N}\big\{\delta(\sigma,\sigma')\big\},\, A_N(\beta,t)\big]\,
\leq\, \E\big[ 1,\, A_N(\beta,t)\big]\,
=\, \P(A_N(\beta,t))\, .
\end{equation*}
So, by Lemma \ref{lem:COM},
\begin{equation}
\label{eq:Helper2}
\frac{1}{N} \ln\E\big[  \Omega_{\beta,N}\big\{\delta(\sigma,\sigma')\big\},\, A_N(\beta,t)\big]\,
\leq\, - \frac{1}{2} t^2 + N^{-1} \ln(2)\, .
\end{equation}
Putting equations (\ref{eq:Helper1}) and (\ref{eq:Helper2}) together
we obtain
\begin{equation}
\label{eq:hola}
\begin{split}
&\frac{1}{N} \ln \E\big[  \Omega_{\beta,N}\big\{\delta(\sigma,\sigma')\big\}\big]\\
&\hspace{25pt} \leq\,
N^{-1} \ln(2) +
\max\Big\{-\frac{1}{2} t^2 + N^{-1} \ln(2)\, ,\
- \frac{1}{4} (\beta_c - \beta_n)^2 + \frac{1}{2} \beta_c [\beta-\beta_n] + \beta t - o(1) \Big\}\, .
\end{split}
\end{equation}
If we now take
\begin{equation*}
\beta  <  \beta_n + \frac{1}{2} \beta_c [1-(\beta_n/\beta_c)]^2
\end{equation*}
then it is clear that by choosing $t$ positive, but small enough, we will
have a strictly negative limsup of the left hand side of (\ref{eq:hola})
as $N \to \infty$.
Choosing any $0<a<1/2$, let us take $g(\beta) = \beta + a \beta_c[1-(\beta/\beta_c)]^2$.
Then for $\beta_{n+1}=g(\beta_n)$, we have proved the induction step: for
$\b$ in the range $[0,\beta_{n+1}]$,
inequality (\ref{ineq:overlap}) also holds.
%
\end{proof}
We complete this section with the proof of Proposition \ref{prop:LowerBd}.

\noindent
\begin{proof}
By Lemma \ref{prop:overlap}, the integrand of the third term in the right
hand side of Eq. (\ref{sumrule-beta}) approaches $0$ uniformly as $N\to\infty$,
as long as $0\leq \beta< \beta_c$. This gives the desired result.
 \end{proof}

\section{The Generalized Random Energy Model}

In this section we extend the method developed in the previous section to treat the GREM.
This is essentially a ``correlated random energy model" on a hierarchical graph -- that is, a tree.

\subsection{Set-up and Basics}

The GREM is a family of models, taking various parameters for the definition. Let $n \in \N_+$ be an integer, equal to the number of levels in the hierarchical tree. Let $K_1,\dots,K_n$ be positive integers
such that $K_1+K_2+\dots+K_n = N$, where $N$ is the system size.
Also, let $a_1,\dots,a_n$ be real numbers such that $0<a_i$ for $i=1,\dots,n$ and $a_1+a_2+\dots+a_n=1$.
\begin{definition}
Given $N \in \mathbf{N}$ and $\s \in \Sigma_N$, for $i=1,\ldots,n$, let $\pi_i(\s)$ be
the canonical projection over the subset $\Sigma_{K_i}$ generated by
the lexicographical partition ${\cal P}$ of the coordinates $(\s_1,\ldots,\s_N)$ into the
first $K_1$ coordinates, the successive $K_2$ coordinates and so on up to the last
$K_n$ coordinates. Namely, $\Sigma_N = \Sigma_{K_1} \times\ldots\times \Sigma_{K_n}$,
$\otimes_{i=1}^n \pi_i = 1_{\Sigma_N}$ and
$\pi_i(\s) = (\s_{K_1+\dots+K_{i-1}+1},\dots,\s_{K_1+\dots+K_{i}})$.
\end{definition}
Then the GREM Hamiltonian is a family of
Gaussian random variables having the covariance
\be
\label{grem:covH}
\av{H_N(\s) H_N(\s')}\, =\,
\frac{N}{2}\sum_{i=1}^n a_i \,\prod_{j=1}^i\delta(\pi_j(\s),\pi_j(\s'))\,
\ee
An explicit form of GREM Hamiltonian is
\be
\label{grem-power2}
H_N(\s) = \sqrt{\frac{N}{2}} \sum_{i=1}^n \sqrt{a_i}\,X(\pi_1(\s),\dots,\pi_i(\s))\, .
\ee
where, for each $i=1,\dots,n$, the family of random variables $\{X(\pi_1(\s),\dots,\pi_i(\s))\}_{\s\in\Sigma_N}$
are $2^{K_1+K_2+\dots +K_i}$ i.i.d.\  Gaussians,
and each
family is independent of the others.
\begin{remark}
The Hamiltonian (\ref{grem-power2}) corresponds to a tree with branching number
that at each level is a power of two. We will stick to this case to
simplify the notation, while the more general case of arbitrary branching number
(with the constraint of having approximately $2^N$ leaves in the last layer) is completely
equivalent in the thermodynamic limit.
In order to make statements that apply in the limit, we will consider sequences of $N$'s
and $K_1,\dots,K_n$ such that there are rational numbers $\kappa_1,\dots,\kappa_n$,
all nonnegative, and summing to 1, with $K_i = N \kappa_i$ for each $i=1,\ldots,n$.
\end{remark}
We now prove the variational expression for the pressure of the GREM.
The strategy is to apply the results obtained for the REM model
in the previous section at each level in the hierarchy.
In order to denote the dependence on the parameters $\boldsymbol{a} = (a_1,\dots,a_n)$ and
$\boldsymbol{\kappa} = (\kappa_1,\dots,\kappa_n)$,
let us write the GREM pressure as
\be
P^{(n)}_N(\beta;\boldsymbol{a},\boldsymbol{\kappa})\, =\,
\frac{1}{N} \, \E \, [\ln Z_N (\beta;\boldsymbol{a},\boldsymbol{\kappa})]\, .
\ee
and its thermodynamic limit as
$P^{(n)}(\beta;\boldsymbol{a},\boldsymbol{\kappa})\, :=\,
\lim_{N\rightarrow\infty} P^{(n)}_N(\beta;\boldsymbol{a},\boldsymbol{\kappa})$.

\subsection{Upper bound}

\begin{proposition}
\label{prop:GREMupper}
Consider the GREM model, for which $C_H(\s,\s')$ is given by (\ref{grem:covH}).
For an index $\a = (\a_1,\a_2,\ldots,\a_n) \in {\cal A}^n$ let the random weights
$w(\a)$ be given by
\be
w(\a_1,\ldots,\a_n) = \exp\left[\frac{y(\a_1)}{m_1}\right]\exp\left[\frac{y(\a_1,\a_2)}{m_2}\right]
\cdots\exp\left[\frac{y(\a_1,\ldots , \a_n)}{m_n}\right]
\ee
where the Poisson point processes is now a cascade with
intensity measure $e^{-y}dy$. Namely, $y(\a_1)$ is the usual PPP, then for each given $\a_1$,
$y(\a_1,\a_2)$ is an independent copy (labelled by $\a_1$) of the PPP,...
and so on up to $y(\a_1,\ldots,\a_n)$ which,
for each given $\a_1, \ldots , \a_{n-1}$, is an independent copy of the PPP
(labelled by $\a_1,\ldots,\a_{n-1}$).
We also choose a sequence $0 < m_1 \leq m_2 \leq \dots \leq m_n < 1$ and
\be
\label{grem:covK}
C_K(\a,\a') = (b-1) \frac{N}{2}\sum_{i=1}^n a_i \;\prod_{j=1}^i\delta(\a_j,\a'_j)
\ee
\be
\label{grem:covV}
C_V(\s, \a, \s', \a') = b \frac{N}{2}\sum_{i=1}^n a_i \;\prod_{j=1}^i \delta(\pi_j(\s),\pi_j(\s'))
\,\delta(\a_j,\a'_j),
\ee
where $b$ is a real number such that $b>1$.
Then we obtain the optimal upper bound:
\be
\label{grem:upper-bound}
P^{(n)}_N(\beta;\boldsymbol{a},\boldsymbol{\kappa}) \le
\inf_{0< m_1\leq \dots\leq m_n < 1}\, \sum_{i=1}^n \Big[\frac{\kappa_i}{m_i} \ln(2) +
\frac{\beta^2}{4} \,m_i a_i\Big]\, .
\ee
\end{proposition}
\begin{proof}
It will be along the lines of the proof of Proposition \ref{prop:upper}.
Lemma \ref{QI} is applicable because for each $i=1,\ldots, n$ one has
\be
b \,\prod_{j=1}^i \delta(\pi_j(\s),\pi_j(\s')) \delta(\a_j,\a'_j) \le
\prod_{j=1}^i \delta(\pi_j(\s),\pi_j(\s'))
+ (b-1) \, \prod_{j=1}^i\delta(\a_j,\a'_j)
\ee
For the first term of Eq.(\ref{upper-ineq}), using Eq. (\ref{grem:covV}), we have
\bea
& & \frac{1}{N}\;\av{\ln \sum_{\s,\;\a}w(\a)\exp [-\b V(\s,\a)]}  =  \\
& & \frac{1}{N}\; \mathbb E\left( \ln \sum_{\s,\;\a}
\exp\left[\frac{y(\a_1)}{m_1}\right]
\exp\left[\frac{y(\a_1,\a_2)}{m_2}\right]
\cdots
\exp\left[\frac{y(\a_1, \cdots,\a_n)}{m_n}\right] \right. \nonumber \\
&&
\qquad\qquad\qquad\qquad
\left.
\exp\left[-\b \sqrt{\frac{b N}{2}}
\sum_{i=1}^n \sqrt{a_i}\, X(\pi_1(\s), \pi_2(\s), \ldots, \pi_i(\s), \a_1 , \a_2,  \ldots ,
\a_i )\right] \right)  \nonumber \\
\nonumber
\eea
Since the sum over configurations $\s\in \Sigma_N,\a\in {\cal A}^n$ can be decomposed into
$n$ sums over each subset $\pi_i(\s) \in \Sigma_{K_i}, \a_i \in {\cal A} $ for $i=1,\ldots,n$,
the invariance property (\ref{invariance})
can now be applied telescopically, starting at the $n$th level and tracing back up
to the first level. After this simplification we obtain
\bea
\label{uno}
& & \frac{1}{N}\;\av{\ln \sum_{\s,\;\a}w(\a)\exp [-\b V(\s,\a)]}  = \\
& & \frac{1}{N}\; \av{\ln \sum_{\a} w(\a)} +
\sum_{i=1}^n \frac{1}{m_i N}\ln \av{Z_{K_i}^{m_i}(\b\sqrt{b a_i})} \leq  \nonumber \\
& & \frac{1}{N}\; \av{\ln \sum_{\a} w(\a)} +
\sum_{i=1}^n \frac{1}{m_i N}\ln \av{Z_{K_i}(m_i \b\sqrt{b a_i})} = \nonumber \\
& & \frac{1}{N}\; \av{\ln \sum_{\a} w(\a)} +
\sum_{i=1}^n \Big[\frac{\kappa_i}{m_i} \ln(2)\Big] +
\frac{\b^2}{4}\; b \sum_{i=1}^n a_i m_i\nonumber
\eea
where in the third line we have made use again of Eq.(\ref{ciao}).
For the second term of Eq.(\ref{upper-ineq}), using Eq.(\ref{grem:covK}) and the
invariance property  (\ref{invariance}) we have
\bea
\label{due}
& & \frac{1}{N}\;\av{\ln {\sum_{\a} w(\a)\exp [-\b K(\a)]}} = \\
& & \frac{1}{N}\;\av{\ln \sum_{\a} w(\a)} +
\b^2 \frac{(b-1)}{4}\sum_{i=1}^n a_i m_i \nonumber
\eea
Putting together Eq.(\ref{uno}) and (\ref{due}) we arrive
at the upper bound stated in the Proposition.
\end{proof}
\begin{remark}
\label{grem:actualvalue}
In the following we make the assumption
\be
\label{cond:nondegen}
\frac{\kappa_1}{a_1 } < \frac{\kappa_2}{a_2} < \dots < \frac{\kappa_n}{a_n }
\ee
in order to have a totally nondegenerate sequence of transition temperatures.
To express the inequality of (\ref{grem:upper-bound}) in a more transparent form it is
convenient to introduce a succession of critical temperatures: for $i=1,\ldots,n$ let
$\beta_i^* = \beta_c \sqrt{\frac{\kappa_i}{a_i}}$
(where $\beta_c=2\sqrt{\ln(2)}$ as in the REM). Under the condition (\ref{cond:nondegen}), this implies $\beta_1^*<\dots<\beta_n^*$.
Because of  the constraint $0<m_1\leq \dots \leq m_n< 1$ the optimal $m_i$ is
\be
m_i\, =\, \min\{1,\beta_i^*/\beta\}
\ee
for $i=1,\dots,n$, the value 1 being attained by taking $m_i \uparrow 1$ in the infimum of
Eq. (\ref{grem:upper-bound}).
Therefore, one has
\begin{equation}
P^{(n)}_N(\beta;\boldsymbol{a},\boldsymbol{\kappa})\, \leq\, \mathcal{Q}^{(n)}(\beta;\boldsymbol{a},\boldsymbol{\kappa})\, ,
\end{equation}
where
\be
\mathcal{Q}^{(n)}(\beta;\boldsymbol{a},\boldsymbol{\kappa})\
\, =\,
\left\{\begin{array}{lll}
\sum_{k=1}^n \frac{1}{4} \, a_k (\beta^2 + (\beta_k^*)^2) & \textrm{ for } \beta < \beta_1^*\\
\sum_{k=1}^i \frac{1}{2} \;a_k \beta \beta_k^*  +
\sum_{k=i+1}^n \frac{1}{4} \, a_k (\beta^2 + (\beta_k^*)^2) & \textrm{ for } \beta_i^* \leq \beta \leq \beta_{i+1}^* \\
\sum_{k=1}^n \frac{1}{2} \;a_k \beta \beta_k^* & \textrm{ for } \beta \geq \beta_n^* \\
\end{array} \right.
\end{equation}
\end{remark}

\subsection{Lower Bound}
%
Let us denote the REM pressure (Eq. (\ref{press})) as $P^{(1)}_N(\beta)$, and its thermodynamic
limit (Eq. (\ref{tdlim-press})) as $P^{(1)}(\beta)$.
This is not really an abuse of notation because if $n=1$ then the GREM is the REM, and $a_1=\kappa_1=1$.
In the same way, we write $\mathcal{Q}^{(1)}(\beta) = \mathcal{Q}(\beta)$, where
$\mathcal{Q}(\beta)$ is defined in Eq. (\ref{Q-rem}).

Then the lower bound is the following:
\begin{proposition}
\label{prop:GREMHi-T}
For all $\beta\geq 0$,
\be
P^{(n)}(\beta;\boldsymbol{a},\boldsymbol{\kappa})\,
\geq\, \mathcal{Q}^{(n)}(\beta;\boldsymbol{a},\boldsymbol{\kappa})\, .
\ee
\end{proposition}
\begin{proof}
The proof will follow if we show that
\begin{equation}
\label{ineq:GREMtoREM}
P^{(n)}_N(\beta;\boldsymbol{a},\boldsymbol{\kappa})\,
\geq\, \sum_{i=1}^n \kappa_i P^{(1)}_{K_i}\big(\sqrt{a_i/\kappa_i}\, \beta\big)\, .
\end{equation}
Indeed, taking the thermodynamic limit $N\rightarrow\infty$ on both sides and using
\be
\mathcal{Q}^{(n)}(\beta;\boldsymbol{a},\boldsymbol{\kappa})\, =\, \sum_{i=1}^n \kappa_i \mathcal{Q}^{(1)}\big(\sqrt{a_i/\kappa_i}\, \beta\big)\quad
\textrm{ for }\quad
\text{for all $\beta\geq 0$.}
\ee
we obtain the Lemma statement. To prove (\ref{ineq:GREMtoREM}) we introduce
the interpolating pressure
\be
\label{last-interpol}
\frac{1}{N} \, \E \ln \sum_{\s\in\Sigma_N} e^{-\beta \tilde H(\sigma, t)} \,
\ee
with
\be
\tilde H(\s, t) =
\sqrt{t} \left[ \sqrt{\frac{N}{2}} \sum_{i=1}^n \sqrt{a_i}\,X(\pi_1(\s),\dots,\pi_i(\s))\right ] +
\sqrt{1-t} \left[ \sqrt{\frac{N}{2}} \sum_{i=1}^n \sqrt{a_i}\,Y(\pi_i(\s))\right ] \,
\ee
where the $X$'s and $Y$'s are families of i.i.d. Gaussian random variables, each
independent from the other. A straightforward differentiation of Eq. (\ref{last-interpol})
combined with integration by parts yields Eq. (\ref{ineq:GREMtoREM}).
\end{proof}

\appendix

\section{Proof of Lemma \ref{lem:COM}: Concentration of measure}

In this Appendix we prove the ``concentration of measure'' inequality
(Lemma \ref{lem:COM}). The proof is a generalization\footnote{
In \cite{T} the generalization of Corollary 2.2.5 to $p$-spin models is implicit:
for example in the proof of Theorem 6.1.2.
But our generalization to all models satisfying ``thermodynamic stability'' is new.}
of the proof for the SK model (Corollary 2.2.5 of \cite{T}).

We show that the standard deviation inequality apply to a large class of Gaussian
spin-glass models, which includes both mean-field (SK,p-spin,REM,GREM) and
finite dimensional models (Edwards-Anderson, Random Field).
We recall the basic result for a function of Gaussian variables:
\begin{theorem}[Talagrand]
\label{thm:TCOM}
Consider a Lipschitz function $F$ on $\R^M$, of Lipschitz constant $A$.
If $J_1,\dots,J_M$ are independent, standard normal random variables, then for each $t>0$,
\be
\P\big\{\big|F(\boldsymbol{J}) - \E[F(\boldsymbol{J})]\big| \geq At\big\}\,
\leq\, 2 e^{-t^2/4}\, .
\ee
\end{theorem}
This is Theorem 2.2.4 of \cite{T}.
Talagrand proves this using the ``smart path" method, which is his adaptation of the quadratic interpolation argument.
Thus, his proof in \cite{T} differs from his earlier proofs \cite{T1}
and from the proofs of others \cite{Pisier}.
This is good for those studying spin glasses.
Particularly, one technique is unifying and simplifying various
tools.
Another application of quadratic interpolation is continuity
of the pressure with respect to the covariance of a spin glass Hamiltonian:
Corollary 3.3 of \cite{AiSS2}.
Also, a result which uses the same idea, and surprisingly predates the applications
in spin glasses, is Slepian's lemma \cite{J-DPP}.

Let us consider the general Hamiltonian given by
\be
\label{general-ham}
H_{\Lambda}(\sigma;\boldsymbol{J})\, =\, -\frac{1}{\sqrt{2}} \sum_{X \in \Lambda} \Delta_X J_X \s_X\, ,
\ee
where $\Lambda \subset \Z^d$, $\s_X = \prod_{i\in X} \s_i$,
$\{\Delta_X\}_{X\in\Lambda} \geq 0$ and the $\{J_X\}_{X\in\Lambda}$ are a family of i.i.d. standard Gaussian
random variables, $\av{J_X} = 0$, $\av{J_X J_Y} = \delta_{X,Y}$.
In order to have a bounded quenched pressure we assume the following thermodynamic stability
condition holds:
there exists a constant $c < \infty$ such that
\be
\label{stability}
\sup_{\Lambda \subset \Z^d} \frac{1}{|\Lambda|}\sum_{X\subset \Lambda} \Delta_X^2 \leq c
\ee
We remark that this condition immediately entails the existence of thermodynamic limit
for short-range models \cite{CGr} and it also implies the validity of the Ghirlanda-Guerra
identities both for short-range and mean-field models \cite{GG,CGi,CG1}.
To prove Lemma \ref{lem:COM}, we need to show that the random pressure function
\be
P_{\Lambda}({\boldsymbol{J}})\, =\, \frac{1}{|\Lambda|} \ln \sum_{\sigma \in \Sigma_N} e^{-\beta H_{\Lambda}(\sigma;\boldsymbol{J})}
\ee
is Lipschitz.
For this, we note that
\be
\label{diff-pressure}
P_{\Lambda}({\boldsymbol{J}}) - P_{\Lambda}({\boldsymbol{J}}') =
\int_{0}^1 \frac{dP_{\Lambda}(t{\boldsymbol{J}} + (1-t){\boldsymbol{J}}')}{dt} \; dt \,.
\ee
On the other hand
\be
\frac{dP_{\Lambda}(t{\boldsymbol{J}} + (1-t){\boldsymbol{J}}')}{dt} =
\frac{\beta}{\sqrt{2}\,|\Lambda|} \sum_{X\subset \Lambda} \omega_t(\s_X) \Delta_X (J_X - J_X')
\ee
with
\be
\omega_t(\s_X) = \frac
{\sum_{\s} \s_X e^{-\beta H_{\Lambda}(\sigma;t{\boldsymbol{J}} + (1-t){\boldsymbol{J}}')}}
{\sum_{\s} e^{-\beta H_{\Lambda}(\sigma;t{\boldsymbol{J}} + (1-t){\boldsymbol{J}}')}}
\ee
From Eq. (\ref{diff-pressure}), by using $|\s_X|\leq 1$, Cauchy-Schwarz inequality
and the thermodynamic stability condition (\ref{stability}), it then follows
\be
|P_{\Lambda}({\boldsymbol{J}}) - P_{\Lambda}({\boldsymbol{J}}')| \leq
\beta \sqrt{\frac{c}{2 |\Lambda|}} \,\|\boldsymbol{J} - \boldsymbol{J}'\|\, .
\ee
where $\|\cdot\|$ denotes the $L_2$-norm.
Therefore, $P_{\Lambda}({\boldsymbol{J}})$ is Lipschitz,
with Lipschitz constant $A = \beta \sqrt{\frac{c}{2 |\Lambda|}}$.
Applying Theorem \ref{thm:TCOM} it gives
\be
\P\big\{\big|P_{\Lambda}(\boldsymbol{J}) - \E[P_{\Lambda}(\boldsymbol{J})]\big| \geq t\big\}\,
\leq\, 2 \exp \left({- \frac{t^2 |\Lambda|}{2 c \beta^2}}\right)\, .
\ee
This result apply equally well to all general Hamiltonian of the
form (\ref{general-ham}). The REM model is obtained from Eq. (\ref{general-ham})
with the choice $|\Lambda| = N$, $\Delta_X = \sqrt{N 2^{-N}}$ (this is indeed Hamiltonian
(\ref{eq:HFunJ})). In this case the condition (\ref{stability}) gives $c=1$ and the
statement of Lemma \ref{lem:COM} is proved.


\end{document}